\documentclass[12pt]{article}

         \usepackage{amsfonts}
         \usepackage{amssymb}

\def\be{\begin{eqnarray}}
\def\ee{\end{eqnarray}}
\def\bee{\begin{eqnarray*}}
\def\eee{\end{eqnarray*}}

\newtheorem{thm}{Theorem}

\newtheorem{lemma}[thm]{Lemma}

\def\Tr{{\rm Tr}}

           \title{
Inequalities for trace norms of $2 \times 2$ block matrices}

         \author{Christopher King  
\\
\\ Department of Mathematics
\\ Northeastern University
\\ Boston MA 02115
\\ 
{\normalsize king@neu.edu}
}
\begin{document}

\maketitle

\begin{abstract}
This paper derives an inequality relating the $p$-norm of a positive $2 \times 2$ block
matrix to the $p$-norm of the $2 \times 2$ matrix obtained by replacing each
block by its $p$-norm. The inequality had been known for integer values of $p$,
so the main contribution here is the extension to all values $p \geq 1$.
In a special case the result reproduces Hanner's inequality.
A weaker inequality which applies also to non-positive matrices is
presented. As an application in quantum information theory,
the inequality is used to obtain some results concerning
maximal $p$-norms of product channels.
\end{abstract}

\pagebreak



\section{Introduction and statement of results}
Quantum information theory has raised some interesting mathematical questions
about completely positive trace preserving maps.
Such maps describe the evolution of open quantum systems, or quantum
systems in the presence of noise \cite{BS}. 
Many of these questions are related to the quantum entropy of states,
and the associated notion of the trace norm, or $p$-norm, of a state.
In one case \cite{K1} the investigation of the additivity question for product channels
(which will be explained in Section 5) led
to an inequality for $p$-norms of positive $2 \times 2$ block matrices for integer values of $p$.
The present paper is devoted to showing that this inequality extends to non-integer values of
$p$. Some implications of this result for the additivity question are
presented, as well as a somewhat weaker inequality which applies to
all $2 \times 2$ block matrices.

The inequality for positive matrices turns out to be closely related to Hanner's inequality \cite{Ha},
which itself relates to the uniform convexity of the matrix spaces $C_p$
(these matrix spaces are the non-commutative versions of the function spaces $L_p$).
The precise relation between these results will be described after the statements of Theorem \ref{thm1}
and Theorem \ref{thm2} below. Hanner's inequality and uniform convexity
for $C_p$ were first established by Tomczak-Jaegermann \cite{T-J} for special values
of $p$, and later proved for all $p \geq 1$ by Ball, Carlen and Lieb \cite{BCL}.
Many of the ideas and methods used in the proofs
of Theorems \ref{thm1} and \ref{thm2} in this paper are taken from 
the paper by Ball, Carlen and Lieb. The heart of the proof of Theorem \ref{thm1} is the convexity
result presented below in Lemma \ref{lemma3}, which extends a result used by Hanner \cite{Ha}
in his original paper.

\medskip
Let $M$ be a $2n \times 2n$ positive semi-definite matrix. It can be written
in the block form
\be\label{def:M}
M = \pmatrix{ X & Y \cr Y^{*} & Z}
\ee
where $X,Y,Z$ are $n \times n$ matrices. The condition $M \geq 0$ requires that 
$X \geq 0$ and $Z \geq 0$, and also that
$Y = X^{1/2} R Z^{1/2}$ where $R$ is a contraction. 

Recall that the $p$-norm of a matrix $A$ is defined as
\be
|| A ||_p = \bigg( \Tr ( A^{*} A)^{p/2} \bigg)^{1/p}
\ee
Define the $2 \times 2$ matrix
\be\label{def:m}
m = \pmatrix{ || X ||_p & || Y ||_p \cr || Y ||_p & || Z ||_p }
\ee
From H\"older's inequality it follows that
\be
|| Y ||_p = || X^{1/2} \, R Z^{1/2} ||_p \leq || X ||_p^{1/2} \,\, || Z ||_p^{1/2}
\ee
which implies that $m \geq 0$ also.

\bigskip

\bigskip
\begin{thm}\label{thm1}
Let $M$ and $m$ be defined as in (\ref{def:M}) and (\ref{def:m}).
The following inequalities hold:
\medskip
\par\noindent {\bf a)} for $1 \leq p \leq 2$,
\be\label{thm1.1}
|| M ||_p \geq || m ||_p
\ee
\par\noindent {\bf b)} for $2 \leq p \leq \infty$,
\be\label{thm1.2}
|| M ||_p \leq || m ||_p
\ee
\end{thm}

\medskip
Theorem \ref{thm1} is easily proved for integer values of $p$ using
H\"older's inequality (see \cite{K1} for details). In the case where
$X=Z$ and $Y=Y^{*}$, the norms of $M$ and $m$ simplify in the following way:
\be
|| M ||_{p}^p & = & || X + Y ||_{p}^p + || X - Y ||_{p}^p \\
|| m ||_{p}^p & = & \bigg( || X ||_p + || Y ||_p \bigg)^p
+ \bigg| || X ||_p - || Y ||_p \bigg|^p
\ee
With these substitutions, the inequalities (\ref{thm1.1}) and (\ref{thm1.2}) 
are seen to be special cases of
Hanner's inequality \cite{Ha} for the matrix spaces $C_p$. As mentioned above,
Hanner's inequality for $C_p$ was proved by Tomczak-Jaegermann \cite{T-J} for special values
of $p$, and later proved for all $p \geq 1$ by Ball, Carlen and Lieb \cite{BCL}.

\medskip
The next Theorem presents a weaker pair of
inequalities which hold for all $2 \times 2$ block matrices.

\bigskip

\bigskip
\begin{thm}\label{thm2}
Let $X$, $Y$, $Z$, $W$ be complex $n \times n$ matrices.
Define the $2 \times 2$ symmetric matrix
\be\label{def:alpha}
\alpha = \pmatrix{||X||_p & \Big( {1 \over 2} ||Y||_p^{p} + {1 \over 2} ||W||_p^{p} \Big)^{1/p} \cr
\Big( {1 \over 2} ||Y||_p^{p} + {1 \over 2} ||W||_p^{p} \Big)^{1/p} & ||Z||_p}
\ee
The following inequalities hold:
\medskip
\par\noindent {\bf a)} for $1 \leq p \leq 2$,
\be\label{thm2.1}
\bigg|\bigg| \pmatrix{X & Y \cr W & Z} \bigg|\bigg|_p \geq 
2^{1/p} \bigg[ {p-1 \over 2}\,\, \Tr (\alpha^2) + {2 - p \over 4}\,\, (\Tr \alpha )^2 \bigg]^{1/2}
\ee
\par\noindent {\bf b)} for $2 \leq p \leq \infty$,
\be\label{thm2.2}
\bigg|\bigg| \pmatrix{X & Y \cr W & Z} \bigg|\bigg|_p \leq 
2^{1/p} \bigg[ {p-1 \over 2}\,\, \Tr (\alpha^2) + {2 - p \over 4}\,\, (\Tr \alpha )^2 \bigg]^{1/2}
\ee
\end{thm}

\medskip
Again considering the special case where $X = X^{*} = Z$ and $Y = Y^{*} = W$, the right side
of (\ref{thm2.1}) and (\ref{thm2.2}) becomes
\be
2^{1/p} \bigg[ ||X||_p^2 + (p-1) \,\, ||Y||_p^2 \bigg]^{1/2}
\ee
The inequalities in this case were derived in \cite{BCL}, and used to establish the
2-uniform convexity (with best constant) of the space $C_p$.
When the block matrix $M$ on the left side 
of (\ref{thm2.1}) is positive and defined as in (\ref{def:M}), the 
inequality can be easily derived from Theorem \ref{thm1}, as follows.
Observe that in this case
\be\label{new.m}
|| m ||_p = \bigg( (u + v)^p + (u - v)^p \bigg)^{1/p}
\ee
where
\be
u & = & {||X||_p + ||Z||_p \over 2} \\
v & = & \Bigg[ \Bigg({||X||_p - ||Z||_p \over 2}\Bigg)^2 + ||Y||_p^2 \Bigg]^{1/2}
\ee
Gross's two-point inequality \cite{G} states that for all
real numbers $a$ and $b$, and all $1 \leq p \leq 2$,
\be\label{Gross}
\bigg( |a+b|^p + |a-b|^p \bigg)^{1/p} \geq 2^{1/p} 
\bigg( a^2 + (p-1) \,b^2 \bigg)^{1/2}
\ee
Applying Gross's inequality to the right side of (\ref{new.m}) and using
(\ref{thm1.1}) immediately
gives (\ref{thm2.1}). In section 3 we prove Theorem \ref{thm2} in the general case 
(where positivity is not assumed) by 
using some very non-trivial results from the paper \cite{BCL}.

\medskip
Most of the new work in this paper goes into the proof of 
Theorem \ref{thm1}, part (a). The proof has three main ingredients: for convenience we state them as
separate lemmas here. The first ingredient is a slight modification  of a convexity result
from \cite{BCL}.

\medskip
\begin{lemma}\label{lemma2}
Let $M = \pmatrix{X & Y \cr Y^{*} & Z} \geq  0$ where $X,Y,Z$ are $n \times n$ matrices.
For fixed $Y$, and for $1 \leq p \leq 2$, the function
\be
(X,Z) \longmapsto \Tr M^p - \Tr X^p - \Tr Z^p
\ee
is jointly convex in $X$ and $Z$.
\end{lemma}

\medskip
The second ingredient extends a convexity result of Hanner \cite{Ha}
to the case of positive $2 \times 2$ matrices
with positive coefficients. 

\medskip
\begin{lemma}\label{lemma3}
Let $A = \pmatrix{a & c \cr c & b} > 0$ where $a,b,c \geq 0$.
For $1 \leq p \leq 2$, the function
\be\label{def:g}
g(A) = \Tr \pmatrix{a^{1/p} & c^{1/p} \cr c^{1/p} & b^{1/p}}^p
\ee
is convex in $A$.
\end{lemma}

\medskip
The third ingredient is a monotonicity result for positive $2 \times 2$ matrices.

\medskip
\begin{lemma}\label{lemma4}
Let $A = \pmatrix{a & c \cr c & b} > 0$ where $a,b,c \geq 0$.
For fixed $c$, and for $1 \leq p \leq 2$, the function
\be\label{def:h}
(a,b) \longmapsto \Tr A^{p} - a^{p} - b^{p}
\ee
is decreasing in $a$ and $b$.
\end{lemma}

\medskip
The paper is organised as follows. In Section 2 we present the proof of Theorem \ref{thm1}
using Lemmas \ref{lemma2}, \ref{lemma3} and \ref{lemma4}. Section 3 contains the proof of
Theorem \ref{thm2}, which is mostly a straightforward adaptation of the proof
of the uniform convexity result in \cite{BCL}.
Lemmas \ref{lemma2}, \ref{lemma3} and \ref{lemma4} are proved in Section 4,
and Section 5 describes an application of Theorem \ref{thm1} in Quantum Information Theory.

\section{Proof of Theorem \ref{thm1}}
Many of the ideas in this proof are taken from the proof of Hanner's inequality
in \cite{BCL}. First, we borrow the duality argument from Section IV of that paper
to show that part (b) follows from part (a). For $p \geq 2$ define $q \leq 2$ to be
its conjugate index. Then there is a $2n \times 2n$ matrix $K$ satisfying
$ || K ||_q = 1$ such that
\be
|| M ||_p = \sup_{L: || L ||_q =1} | \, \Tr ( L M ) \, | = \Tr ( K M )
\ee
The positivity of $M$ means that $K$ can be assumed to be positive. 
Let 
\be
K = \pmatrix{A & C \cr C^{*} & B} \geq 0
\ee
then
\be\label{a->b}
\Tr ( K M ) & = & \Tr (A X) + \Tr ( C Y^{*} ) + \Tr (C^{*} Y ) + \Tr ( B Z ) \\ \nonumber
& \leq & || A ||_q \, || X||_p + 2 ||C||_q \, ||Y||_p + ||B||_q \, ||Z||_p \\ \nonumber
& = & \Tr \pmatrix{||A||_q & ||C||_q \cr ||C||_q & ||B||_q} m \\ \nonumber
& \leq & \bigg|\bigg| \pmatrix{||A||_q & ||C||_q \cr ||C||_q & ||B||_q} \bigg|\bigg|_q \, ||m||_p \\
\nonumber & \leq & ||K||_q \, ||m||_p \\ \nonumber
& = & ||m||_p
\ee
The first and second inequalities are applications of H\"older's inequality,
the last inequality uses part (a) of Theorem \ref{thm1}.

\medskip
Next we turn to the proof of part (a) of Theorem \ref{thm1}.
The inequality becomes an equality at the
values $p=1,2$, 
so we will assume henceforth that $1 < p < 2$. Using the singular value
decomposition we can write
\be
Y = U D V^{*}
\ee
where $U,V$ are unitary matrices and $D \geq 0$ is diagonal. Unitary invariance of the
$p$ norm implies that
\be
||M||_p = \bigg|\bigg| \pmatrix{U^{*} X U & D \cr D & V^{*} Z V} \bigg|\bigg|_p
\ee
and also that $||X||_p = ||U^{*} X U||_p$,
$||Z||_p = ||V^{*} Z V||_p$ and $||Y||_p = ||D||_p$.
So without loss of generality we will assume henceforth that $Y$ is diagonal and
non-negative. 

\medskip
Next we use a diagonalization argument from Section III of \cite{BCL}.
Let $U_1, \dots, U_{2^n}$ denote the $2^n$ diagonal $n \times n$
matrices with diagonal entries $\pm 1$. Then for any
$n \times n$ matrix $A$ we have
\be
A_d = \sum_{i=1}^{2^n} 2^{-n} \,\, U_i A U_{i}^{*}
\ee
where $A_d$ is the diagonal part of $A$. Since $Y$ is diagonal this implies that
\be\label{av1}
\sum_{i=1}^{2^n} 2^{-n} \pmatrix{U_i & 0 \cr 0 & U_i} \pmatrix{X & Y \cr Y & Z}
\pmatrix{U_{i}^{*} & 0 \cr 0 & U_{i}^{*}} 
= \pmatrix{X_d & Y \cr Y & Z_d}
\ee
and by the same reasoning
\be\label{av2}
\sum_{i=1}^{2^n} 2^{-n} \pmatrix{U_i & 0 \cr 0 & U_i} \pmatrix{X & 0 \cr 0 & Z}
\pmatrix{U_{i}^{*} & 0 \cr 0 & U_{i}^{*}} 
= \pmatrix{X_d & 0 \cr 0 & Z_d}
\ee

\medskip
Now we combine (\ref{av1}) and (\ref{av2}) with the convexity result Lemma \ref{lemma2},
which gives
\be\label{ineq1}
\Tr \pmatrix{X & Y \cr Y & Z}^p - \Tr \pmatrix{X & 0 \cr 0 & Z}^p 
 \geq 
\Tr \pmatrix{X_d & Y \cr Y & Z_d}^p - \Tr \pmatrix{X_d & 0 \cr 0 & Z_d}^p 
\ee

\medskip
The matrices $X_d,Y,Z_d$ are all diagonal with non-negative entries. Denote these
entries by $(x_1, \dots, x_n)$, $(y_1, \dots, y_n)$ and $(z_1, \dots, z_n)$
respectively. Then 
\be\label{eqn1}
\Tr \pmatrix{X_d & Y \cr Y & Z_d}^p = 
\sum_{i=1}^n \Tr \pmatrix{x_i & y_i \cr y_i & z_i}^p
\ee

\medskip
Now for $i=1,\dots,n$ define
\be
a_i = x_i^p, \quad b_i = z_i^p, \quad c_i = y_i^p
\ee
and introduce the $2 \times 2$ matrices
\be
A_i = \pmatrix{a_i & c_i \cr c_i & b_i}
\ee
It follows that
\be\label{p-norms}
||X_d||_p & = & (a_1 + \cdots + a_n)^{1/p} \\ \nonumber
||Y||_p & = & (c_1 + \cdots + c_n)^{1/p} \\ \nonumber
||Z_d||_p & = & (b_1 + \cdots + b_n)^{1/p}
\ee
and the definition (\ref{def:g}) implies that
\be\label{eqn3}
\Tr \pmatrix{||X_d||_p & ||Y||_p \cr ||Y||_p & ||Z_d||_p}^p
= g(A_1 + \cdots + A_n)
\ee
Furthermore (\ref{eqn1}) implies that
\be\label{eqn2}
\Tr \pmatrix{X_d & Y \cr Y & Z_d}^p =
g(A_1) + \cdots + g(A_n)
\ee
Also, for any positive number $k$ we have $g(k A) = k g(A)$. Combining this with the
convexity result Lemma \ref{lemma3} gives
\be
g(A_1 + \cdots + A_n) \leq g(A_1) + \cdots + g(A_n),
\ee
which from (\ref{eqn2}) and (\ref{eqn3}) implies that
\be\label{ineq2}
\Tr \pmatrix{X_d & Y \cr Y & Z_d}^p \geq 
\Tr \pmatrix{||X_d||_p & ||Y||_p \cr ||Y||_p & ||Z_d||_p}^p
\ee

\medskip
Combining (\ref{ineq1}) with (\ref{ineq2}) gives
\be\label{ineq3}
& \Tr & \pmatrix{X & Y \cr Y & Z}^p - \Tr \pmatrix{X & 0 \cr 0 & Z}^p \\ \nonumber
\geq
& \Tr & \pmatrix{||X_d||_p & ||Y||_p \cr ||Y||_p & ||Z_d||_p}^p -
\Tr \pmatrix{||X_d||_p & 0 \cr 0 & ||Z_d||_p}^p 
\ee
Furthermore
\be
||X_d||_p \leq ||X||_p, \quad\quad
||Z_d||_p \leq ||Z||_p
\ee
Applying Lemma \ref{lemma4} to the right side of (\ref{ineq3}) shows that
\be\label{ineq4}
 & \Tr & \pmatrix{||X_d||_p & ||Y||_p \cr ||Y||_p & ||Z_d||_p}^p -
\Tr \pmatrix{||X_d||_p & 0 \cr 0 & ||Z_d||_p}^p \\ \nonumber
\geq & \Tr & \pmatrix{||X||_p & ||Y||_p \cr ||Y||_p & ||Z||_p}^p -
\Tr \pmatrix{||X||_p & 0 \cr 0 & ||Z||_p}^p 
\ee
Furthermore
\be
\Tr \pmatrix{X & 0 \cr 0 & Z}^p = 
\Tr \pmatrix{||X||_p & 0 \cr 0 & ||Z||_p}^p
\ee
and therefore (\ref{ineq3}) and (\ref{ineq4})
imply the result Theorem \ref{thm1}.

\section{Proof of Theorem \ref{thm2}}
This proof follows very closely the methods in Section III of \cite{BCL}.
First we use a duality argument to deduce (\ref{thm2.2}) from (\ref{thm2.1}).
Let $p \geq 2$ and let $q$ be the index conjugate to $p$. Then
it follows as in (\ref{a->b}) that there is a matrix $K = \pmatrix{
A & C \cr D & B}$ such that $||K||_q =1$ and
\be\label{3.1}
\bigg|\bigg| \pmatrix{X & Y \cr W & Z} \bigg|\bigg|_p & = &
\Tr \, K \, \pmatrix{X & Y \cr W & Z} \\ \nonumber 
& = & \Tr \bigg( AX + CW + DY + BZ \bigg)
\ee
Define
\be
a = ||A||_q, \quad b = ||B||_q, \quad c = \Big({1 \over 2}||C||_q^q + 
{1 \over 2}||D||_q^q \Big)^{1/q}
\ee
and similarly
\be\label{def:x}
x = ||X||_p, \quad z = ||Z||_p, \quad y = \Big({1 \over 2}||Y||_p^p + 
{1 \over 2}||W||_p^p \Big)^{1/p}
\ee
Then applying H\"older's inequality to (\ref{3.1}) gives
\be\label{3.2}
\bigg|\bigg| \pmatrix{X & Y \cr W & Z} \bigg|\bigg|_p \leq
ax + bz + 2 c y
\ee
This is rewritten as
\be\label{3.3}
ax + bz + 2 c y  & = & 2  \Big({a+b \over 2}\Big) \Big({x+z \over 2}\Big)
+ 2 \Big({a-b \over 2}\Big) \Big({x-z \over 2}\Big)  + 2 c y \\ \nonumber
& = & 2 \Big({a+b \over 2}\Big) \Big({x+z \over 2}\Big)  \\ \nonumber
& + & 2 \Big({ q-1}\Big)^{1/2} 
\Big({a-b \over 2}\Big) \Big({1 \over q-1}\Big)^{1/2} \Big({x-z \over
2}\Big)  \\ \nonumber
& + & 2  \Big({q-1}\Big)^{1/2} c \Big({1 \over q-1}\Big)^{1/2} y
\ee
Now we apply the Cauchy-Schwarz inequality to the right side
of (\ref{3.3}); the result is
\be\label{3.3a}
ax + bz + 2 c y 
& \leq & 2 \, \bigg[ \Big({a+b \over 2}\Big)^2 + (q-1) \Big({a-b \over 2}\Big)^2 + (q-1) c^2 \bigg]^{1/2}
\nonumber \\
& \times & \bigg[ \Big({x+z \over 2}\Big)^2 + {1 \over q-1} \Big({x-z \over 2}\Big)^2 + {1 \over q-1} y^2
\bigg]^{1/2}
\ee
Furthermore,
\be
\Big({a+b \over 2}\Big)^2 + (q-1) \Big({a-b \over 2}\Big)^2 + (q-1) c^2
= {q-1 \over 2}\,\, \Tr (k^2) + {2 - q \over 4}\,\, (\Tr k )^2
\ee
where $k$ is the $2 \times 2$ matrix
\be
k = \pmatrix{a & c \cr
c & b}
\ee
Since $q \leq 2$, (\ref{thm2.1}) implies that
\be\label{3.3b}
\bigg[ {q-1 \over 2}\,\, \Tr (k^2) + {2 - q \over 4}\,\, (\Tr k )^2 \bigg]^{1/2} & \leq &
2^{-1/q} \,\, \bigg|\bigg| \pmatrix{A & C \cr D & B} \bigg|\bigg|_q \nonumber \\
& = & 2^{-1/q} \, || K ||_q \nonumber \\
& = & 2^{-1/q}
\ee
Combining (\ref{3.2}), (\ref{3.3a}) and (\ref{3.3b}) gives
\be
\bigg|\bigg| \pmatrix{X & Y \cr W & Z} \bigg|\bigg|_p & \leq &
2^{1-1/q} \,\, \bigg[ \Big({x+z \over 2}\Big)^2 + {1 \over q-1} \Big({x-z \over 2}\Big)^2 + 
{1 \over q-1} y^2 \bigg]^{1/2} \nonumber
\\ & = &
2^{1/p} \,\, \bigg[ \Big({x+z \over 2}\Big)^2 + (p-1) \Big({x-z \over 2}\Big)^2 + (p-1) y^2 \bigg]^{1/2}
\nonumber  \\
& = & 2^{1/p} \bigg[ {p-1 \over 2}\,\, \Tr (\alpha^2) + {2 - p \over 4}\,\, (\Tr \alpha )^2 \bigg]^{1/2}
\ee
where $\alpha$ was defined in (\ref{def:alpha}), and this proves (\ref{thm2.2}).

\medskip
Suppose now that $1 \leq p \leq 2$.
The first step in the proof of (\ref{thm2.1}) is to reduce the result to the case where the matrix is
self-adjoint. This is done by modifying an argument from section III of
\cite{BCL}. Given $X$, $Y$, $W$ and $Z$ define the matrices
\be
J = \pmatrix{X & Y \cr W & Z}
\ee 
and
\be
L = \pmatrix{0 & X & 0 & Y \cr X^{*} & 0 & W^{*} & 0 \cr
0 & W & 0 & Z \cr Y^{*} & 0 & Z^{*} & 0}
\ee
Then $L = L^{*}$ and furthermore
\be\label{3.4}
\Tr | L |^p = \Tr (L^{*} L)^{p/2} = \Tr (J^{*} J)^{p/2} + \Tr (J J^{*})^{p/2} =
2 \, \Tr |J|^p
\ee
Assuming that (\ref{thm2.1}) holds for self-adjoint matrices, it implies that
\be\label{3.5}
|| L ||_p \geq 2^{1/p} \,\,
\bigg[ {p-1 \over 2}\,\, \Tr (\beta^2) + {2 - p \over 4}\,\, (\Tr \beta )^2 \bigg]^{1/2}
\ee
where $\beta$ is given by
\be\label{3.6}
\beta = \pmatrix{2^{1/p} \,||X||_p & \Big( ||Y||_p^{p} + ||W||_p^{p} \Big)^{1/p} \cr
\Big( ||Y||_p^{p} + ||W||_p^{p} \Big)^{1/p} & 2^{1/p} \,||Z||_p}
\ee
Comparing with (\ref{def:alpha}) shows that $\beta = 2^{1/p} \alpha$, and hence
(\ref{3.4}) and (\ref{3.5}) imply (\ref{thm2.1}).

\medskip
The self-adjoint case will be handled by modifying slightly a very non-trivial
proof in section III of the paper \cite{BCL}. 
For convenience we state the hard part of the proof in \cite{BCL} as a separate lemma
here, and refer the reader to the original source for its proof.

\medskip
\begin{lemma}\label{lemma5} {\rm [Ball, Carlen and Lieb]}
Let $A$ and $B$ be self-adjoint $n \times n$ matrices,
with $A$ non-singular, and suppose that $1 \leq p \leq 2$.
Then
\be
{d^2 \over d r^2} \bigg( \Tr |A + r B|^p \bigg)^{2/p} \bigg|_{r=0}
\geq 2 (p-1) \bigg( \Tr |B|^p \bigg)^{2/p}
\ee
\end{lemma}

\medskip
Now suppose that $X$, $Y$ and $Z$ are $n \times n$ complex matrices with
$X$ and $Z$ self-adjoint. Define
\be
F = \pmatrix{X & 0 \cr 0 & Z}, \quad 
G = \pmatrix{0 & Y \cr Y^{*} & 0}
\ee
Using the notation introduced in (\ref{def:x}), the goal is to show that
\be\label{3.7}
\bigg( \Tr |F + r G|^{p} \bigg)^{2/p} \geq 2^{2/p} \, 
\bigg[ \Big({x+z \over 2}\Big)^2 + (p-1) \Big({x-z \over 2}\Big)^2 + (p-1) r^2 y^2 \bigg]
\ee
at the value $r=1$, where now $y = ||Y||_p$. First, it is easy to show
that (\ref{3.7}) holds at $r=0$: in this case the left side is
$(x^p + z^p)^{2/p}$, and Gross's two-point inequality (\ref{Gross}) implies that
\be
(x^p + z^p)^{2/p} \geq 2^{2/p} \Big[ \Big({x+z \over 2}\Big)^2 + 
(p-1) \Big({x-z \over 2}\Big)^2
\Big]
\ee
Second, both sides of (\ref{3.7}) are even functions of $r$ (the left side because 
the matrices $F + r G$ and $F - r G$ have the same spectrum), hence the
derivatives of both sides vanish at $r=0$.
Therefore it is sufficient to prove that 
\be\label{3.8}
{d^2 \over d r^2} \bigg( \Tr |F + r G|^{p} \bigg)^{2/p} \geq 2^{2/p} \, 
2 (p-1) y^2 =  2 (p-1) \bigg( \Tr |G|^p \bigg)^{2/p}
\ee
for all $0 \leq r \leq 1$. The inequality (\ref{3.8}) is established by
the following argument (again borrowed from \cite{BCL}).
By continuity, it can be assumed that the ranges
of $F$ and $G$ span all of ${\bf C}^{2n}$ (recall that $X$, $Y$, $Z$ are $n \times n$
matrices) and therefore that $F + r G$ is non-singular at all but possibly
$2n$ values of $r$ in the interval $0 \leq r \leq 1$. By continuity again it
is sufficient to establish (\ref{3.8}) at these non-singular values.
Let $r_0$ be such a non-singular value, and let $A = F + r_0 G$ and $B = G$.
Then at $r=r_0$, (\ref{3.8}) becomes
\be\label{3.9}
{d^2 \over d r^2} \bigg( \Tr |A + r B|^p \bigg)^{2/p} \bigg|_{r=0}
\geq 2 (p-1) \bigg( \Tr |B|^p \bigg)^{2/p}
\ee
But this is exactly the statement of Lemma \ref{lemma5}, hence (\ref{thm2.1}) is
proved.

\section{Proofs of Lemmas}
\subsection{Proof of Lemma \ref{lemma2}}
This result is a slight modification of a convexity result proved
in Section IV of \cite{BCL}. For a positive matrix $M = \pmatrix{X & Y \cr Y^{*} & Z} \geq  0$,
define  $M_d = \pmatrix{X & 0 \cr 0 & Z} \geq  0$
and $F = M - M_d$. Let
\be
D = \pmatrix{D_1 & 0 \cr 0 & D_2} = D^{*}
\ee
be a block diagonal self-adjoint matrix, and define
\bee
\phi(s) & = & \Tr (M + s D)^p - \Tr (M_d + s D)^p \\
& = & \Tr (M_d + F + s D)^p - \Tr (M_d + s D)^p
\eee
Then for $1 \leq p \leq 2$ the second derivative of $\phi$ has the following integral
representation (see \cite{BCL} for details):
\be\label{phi-der}
{\phi}''(0) & = & p {\gamma}_p \int_{0}^{\infty}
t^{p-1} \Tr \bigg( {1 \over t + M_d + F} D {1 \over t + M_d + F} D - 
{1 \over t + M_d} D {1 \over t + M_d} D  \bigg)  d t \nonumber \\
&& 
\ee
for some constant $\gamma_p$.
Furthermore, the matrices $M_d + F + s D$ and $M_d - F + s D$ have the
same spectrum, hence (\ref{phi-der}) can be written
\be\label{phi-der2}
{\phi}''(0) = {p \over 2} {\gamma}_p \int_{0}^{\infty}
t^{p-1} \Tr \bigg( {1 \over t + M_d + F}  & D & {1 \over t + M_d + F} \,\, D \\ \nonumber
+ {1 \over t + M_d - F} & D & {1 \over t + M_d - F}\,\, D \\ \nonumber
-  2 \,\,{1 \over t + M_d} & D & {1 \over t + M_d}\,\, D \bigg) d t
\ee
Ball, Carlen and Lieb \cite{BCL} proved that for $t \geq 0$, and
for any  self-adjoint matrix $A$, the map
\be
X \longmapsto \Tr {1 \over t + X} A {1 \over t + X} A 
\ee
is convex on the set of positive
matrices. Applying this to (\ref{phi-der2}) with $X = M_d$ and $A = D$
shows that ${\phi}''(0) \geq 0$, which is the convexity result
in Lemma \ref{lemma2}.

\subsection{Proof of Lemma \ref{lemma3}}
Since $g$ is homogeneous it is sufficient to prove that 
\be
g(A + B) \leq g(A) + g(B)
\ee
for any $A,B$ of the specified form. To prove this,
it is sufficient to show that
\be
{d \over dt} g(A + t B) |_{t=0} \leq g(B)
\ee
for any $A,B$. Let
\be
A = \pmatrix{a & c \cr c & b},\quad
B = \pmatrix{x & y \cr y & z}
\ee
Define
\be
M = \pmatrix{a^{1/p} & c^{1/p} \cr c^{1/p} & b^{1/p}},\quad
L = \pmatrix{a^{(1-p)/p} x & c^{(1-p)/p} y \cr
c^{(1-p)/p} y & b^{(p-1)/p} z}
\ee
Then
\be\label{der1}
{d \over dt} g(A + t B) |_{t=0} = \Tr M^{p-1} \, L
\ee
\medskip

The idea of the proof is to maximise the right side of (\ref{der1}) as 
a function of $M$, and show that the maximum is achieved when
$A$ and $B$ are proportional, in which case the bound is an equality.
This will be done by 
explicitly finding the  critical points of $\Tr M^{p-1} \, L$.

\medskip
To this end write the spectral decomposition of $M$ in the form
\be
M = \pmatrix{a^{1/p} & c^{1/p} \cr c^{1/p} & b^{1/p}} = \lambda P_1 + \mu P_2
\ee
where $P_i$ are projectors onto the normalised eigenvectors of $M$,
and $\lambda, \mu$ are the eigenvalues (notice that the
positivity of $A$ and $B$ implies that both $M$ and $L$ are also
positive). If we assume that
$\lambda \geq \mu$ then for some $0 \leq t \leq 1$ we have
\be
a^{1/p} & = & \lambda t + \mu (1-t) \\
c^{1/p} & = &  \sqrt{t(1-t)}(\lambda - \mu) \\
b^{1/p} & = & \lambda (1-t) + \mu t
\ee
Furthermore it also follows that
\be
M^{p-1} = \pmatrix{k_{11} & k_{12} \cr k_{12} & k_{22}} = {\lambda}^{p-1} P_1 + 
{\mu}^{p-1} P_2
\ee
where
\be
k_{11} & = & {\lambda}^{p-1} t + {\mu}^{p-1} (1-t) \\
k_{12} & = &  \sqrt{t(1-t)}({\lambda}^{p-1} - {\mu}^{p-1}) \\
k_{22} & = & {\lambda}^{p-1} (1-t) + {\mu}^{p-1} t
\ee
Substituting into (\ref{der1}) gives
\be\label{der2}
\Tr M^{p-1} \, L = k_{11} a^{(1-p)/p} x + 2 k_{12} c^{(1-p)/p} y
+ k_{22} b^{(p-1)/p} z
\ee
Equation (\ref{der2}) is invariant under a rescaling of $M$.
Define
\be
h = {\mu \over \lambda}, \quad\quad 0 \leq h \leq 1
\ee
then (\ref{der2}) is a function of $t$ and $h$, and can be written as
\be
\Tr M^{p-1} \, L = F(t,h) = F_1 (t,h) x + F_2 (t,h) y + F_3 (t,h) z
\ee
where
\be
F_1(t,h) & = & {t + (1-t) h^{p-1} \over (t + (1-t) h)^{p-1}} \\
F_2(t,h) & = & 2 \bigg(t(1-t)\bigg)^{1 - p/2} {1 - h^{p-1} \over
(1-h)^{p-1}} \\
F_3(t,h) & = & F_1(1-t,h)
\ee

The goal is to maximise $F(t,h)$ over $t$ and $h$. Define
\be
G & = & \Big(t + (1-t) h\Big) \Big(1 - h^{p-1}\Big) - (p-1) (1-h) 
\Big(t + (1-t) h^{p-1}\Big) \\
H & = & \Big((1-t) + t h\Big) \Big(1 - h^{p-1}\Big) - (p-1) (1-h) 
\Big((1-t) + t h^{p-1}\Big)
\ee
and also let
\be
\xi & = & x \Big(t + (1-t) h\Big)^{-p} \\
\eta & = & y (1-h)^{-p} \, \Big(t(1-t)\Big)^{-p/2} \\
\zeta & = & z \Big(1-t + t h\Big)^{-p}
\ee

Then explicit calculation shows that
\be
{\partial F \over \partial t} = G \xi  - (G-H) \eta - H \zeta
\ee
and
\be
{\partial F \over \partial h} = - t(1-t) (p-1) (1 - h^{p-2}) 
(\xi  - 2 \eta + \zeta)
\ee

The critical equations are
\be\label{crit}
{\partial F \over \partial t} = {\partial F \over \partial h} = 0
\ee
One obvious set of solutions is obtained when $t=0$ or $t=1$, or $h=1$.
In all of these cases, the matrix $M$ must be
diagonal, in which case (\ref{der1}) implies
\be
\Tr M^{p-1} \, L = \Tr B = \Tr \pmatrix{x^{1/p} & 0 \cr
0 & z^{1/p}}^p \leq g(B)
\ee
and this establishes the result. If $0 < t < 1$ and $h < 1$, the critical
equations can be written
\be\label{crit2}
G (\xi - \eta) & = & H(\zeta - \eta) \nonumber \\ 
 \xi  -  \eta & = & - (\zeta - \eta)
\ee
It is easy to show that $h < 1$ implies that $G >0$ and $H > 0$,
hence the solution of (\ref{crit2}) satisfies $\xi = \eta = \zeta$.
In this case $M$ must be proportional
to the matrix
\be
\pmatrix{x^{1/p} & y^{1/p} \cr y^{1/p} & z^{1/p}}
\ee
and substituting into (\ref{der1}) then gives
\be
\Tr M^{p-1} \, L = g(B)
\ee
which proves  the result.

\subsection{Proof of Lemma \ref{lemma4}}
By the convexity result Lemma \ref{lemma3}, it is sufficient to prove that
the function $(a,b) \mapsto \Tr A^{p} - a^{p} - b^{p}$ is decreasing as $a, b \rightarrow \infty$.
For $a >> 1$, and for $1 < p < 2$, easy estimates show that
\be
\Tr A^{p} - a^{p} - b^{p} \simeq p c^2 a^{p-2}
\ee
which is indeed decreasing. Similarly for $b$.

\section{Application to qubit maps}
Quantum information theory has generated an interesting conjecture concerning
completely positive maps on matrix algebras.
Let $\Phi$ be a completely positive trace-preserving (CPTP) map
on the algebra of $n \times n$ matrices.
The minimal entropy of $\Phi$ is defined by
\be\label{def:Smin}
S_{\rm min}(\Phi) = \inf_{\rho} S(\Phi(\rho))
\ee
where $S$ is the von Neumann entropy and the $\inf$ runs over $n \times n$
density matrices (satisfying $\rho \geq 0$ and $\Tr \rho = 1$). 
Minimal entropy is conjectured to be additive for product maps, that is,
it is conjectured that
\be\label{conj1}
S_{\rm min}(\Phi_1 \otimes \Phi_2) = S_{\rm min}(\Phi_1) + S_{\rm min}(\Phi_2)
\ee
for any pair of CPTP maps $\Phi_1$ and $\Phi_2$. The conjecture (\ref{conj1}) has been
established in some special cases \cite{S}, \cite{K2}  but a general proof remains elusive.

\medskip
For related reasons,
Amosov, Holevo and Werner \cite{AHW} defined the maximal $p$-norm for a CPTP map to be
\be\label{def:nu}
{\nu}_{p}(\Phi) = \sup_{\rho} || \Phi(\rho) ||_p
\ee
where the $\sup$ runs again over density matrices. They conjectured that this quantity
is multiplicative for product maps, that is
\be\label{AHW}
{\nu}_{p}(\Phi_1 \otimes \Phi_2) = {\nu}_{p}(\Phi_1) \,\, {\nu}_{p}(\Phi_2)
\ee
Holevo and Werner later discovered a family of counterexamples to this conjecture
for $p \geq 4.79$,
using maps which act on $3 \times 3$ or higher
dimensional matrices \cite{WH}. The conjecture remains open 
if at least one of the pair is a qubit map
(which acts on $2 \times 2$ matrices) or if $p \leq 4$. 

\medskip
As an application of Theorem \ref{thm1}, we now show that it implies the
result (\ref{AHW}) in one special case, namely when $\Phi_1$ is the qubit
depolarizing channel and $p \geq 2$. This result was
derived previously using a lengthier argument \cite{K2}, and the purpose of
this presentation is to explore an alternative method which may allow
new approaches to the additivity problem. Indeed,
the method shown below can be easily extended to cover all unital qubit channels
and even some non-unital qubit maps, thus extending the results in \cite{K1}
which were derived for integer values of $p$.
Unfortunately, the restriction to $p \geq 2$ does not allow any conclusions to be drawn about
additivity of minimal entropy.

\medskip
The depolarizing channel $\Delta$ acts on a state 
$\rho = \pmatrix{a & c  \cr \overline{c} & b}$ by
\be
\Delta(\rho) = \lambda \rho + {1 - \lambda \over 2} I =
\pmatrix{{\lambda}_{+} a + {\lambda}_{-} b &  \lambda c \cr 
\lambda \overline{c} & {\lambda}_{-} a + {\lambda}_{+} b}
\ee
where $\lambda$ is a real parameter and
${\lambda}_{\pm} = (1 \pm \lambda)/2$. 
We will suppose here that $0 \leq \lambda \leq 1$. 
The maximal $p$-norm of $\Delta$ is easily computed to be
\be
{\nu}_{p}(\Delta) = \Bigg( \Big({1 + \lambda \over 2}\Big)^p +
\Big({1 - \lambda \over 2}\Big)^p \Bigg)^{1/p}
\ee
Now consider a  positive $2n \times 2n$ matrix $M$:
\be
M =  \pmatrix{A & C \cr C^{*} & B}
\ee
The map $\Delta \otimes I$ acts on $M$ via
\be
(\Delta \otimes I) (M) = 
\pmatrix{{\lambda}_{+} A + {\lambda}_{-} B &  \lambda C \cr 
\lambda C^{*} & {\lambda}_{-} A + {\lambda}_{+} B}
\ee

\medskip
Let $p \geq 2$, and
let $q \leq 2$ be the index conjugate to $p$. Then
as explained at the start of section 2, 
there is a positive $2n \times 2n$ matrix $K$ satisfying
$|| K ||_q = 1$ such that
\be\label{eqn5}
|| (\Delta \otimes I) (M) ||_p = \Tr \bigg(K (\Delta \otimes I) (M) \bigg)
\ee

\medskip
Following the methods used in (\ref{a->b}), this leads to
\be
\Tr \bigg(K (\Delta \otimes I) (M) \bigg) & \leq &
\bigg|\bigg| \pmatrix{
{\lambda}_{+} ||A||_p + {\lambda}_{-} ||B||_p &  \lambda ||C||_p \cr 
\lambda ||C||_p & {\lambda}_{-} ||A||_p + {\lambda}_{+} ||B||_p}
\bigg|\bigg|_{p} \nonumber \\ 
& = & || \Delta(m) ||_p
\ee
where $m$ is the $2 \times 2$ matrix
\be
m = \pmatrix{||A||_p &  ||C||_p \cr 
||C||_p &  ||B||_p}
\ee
By definition of the $p$-norm this implies
\be\label{nextineq}
|| (\Delta \otimes I) (M) ||_p \leq {\nu}_{p}(\Delta) \,\, \Big( ||A||_p + ||B||_p \Big)
\ee

\medskip
Now let $\rho$ be a $2n \times 2n$ density matrix,
\be
\rho = \pmatrix{\rho_{11} & \rho_{12} \cr \rho_{21} & \rho_{22}}
\ee
and consider the case where $M = (I \otimes \Phi)(\rho)$ and $\Phi$ is some other
channel, so that $(\Delta \otimes I) (M) = (\Delta \otimes \Phi)(\rho)$.  Then
\be
A = \Phi(\rho_{11}), \quad  B = \Phi(\rho_{22})
\ee
and hence
\be
||A||_p + ||B||_p \leq \nu_{p}(\Phi) \,\, \Tr (\rho_{11} + \rho_{22})
= \nu_{p}(\Phi)
\ee
Therefore (\ref{nextineq}) implies that
\be\label{ineq5}
|| (\Delta \otimes \Phi) (\rho) ||_p \leq {\nu}_{p}(\Delta) \, \nu_{p}(\Phi)
\ee
Since (\ref{ineq5}) is valid for all $\rho$, we get
\be
{\nu}_{p}(\Delta \otimes \Phi) \leq 
{\nu}_{p}(\Delta) \, \nu_{p}(\Phi)
\ee
and this establishes the result (\ref{AHW}), since the inequality in the other direction
follows by restricting to product states.

\bigskip
{\bf Acknowledgements}
This work was supported in part by
National Science Foundation Grant DMS--0101205.

{~~}

\end{document}